\documentclass[epj]{svjour}
\usepackage{amssymb}
\usepackage{amsmath}
\usepackage{graphicx}

\newcommand{\ie}[0] {{\it i.e.},\ }
\newcommand{\eg}[0] {{\it e.g.},\ }

\newcommand{\KT}[0] {k_{\rm B}T}

\newcommand{\Z}[0]{{\cal Z}}

\newcommand{\Fig}[1] {Figure \ref{#1}}
\newcommand{\Eq}[1] {equation (\ref{#1})}
\newcommand{\Eqs}[2] {equations (\ref{#1}) and (\ref{#2})}
\newcommand{\Sec}[1] {section \ref{#1}}

\newcommand{\A}[0]{\langle A \rangle}
\newcommand{\Amax}[0]{{A_{\rm max}}}
\newcommand{\Pc}[0]{p_{\rm c}}
\newcommand{\Gc}[0]{\gamma_{\rm c}}
\newcommand{\R}[0]{{\bf r}}
\newcommand{\Q}[0]{\hat Q}

\DeclareTextSymbolDefault\textctz{T3}

\begin{document}

\title{Swelling of two-dimensional polymer rings by trapped particles}

\author{Emir Haleva \and Haim Diamant}

\institute{School of Chemistry, Raymond \& Beverly Sackler Faculty of
  Exact Sciences, Tel Aviv University, Tel Aviv 69978, Israel}

\date{Received: date / Revised version: date}

\abstract {The mean area of a two-dimensional Gaussian ring of $N$
  monomers is known to diverge when the ring is subject to a critical
  pressure differential, $\Pc\sim N^{-1}$. In a recent publication
  [Eur.\ Phys.\ J.\ E {\bf 19}, 461 (2006)] we have shown that for an
  inextensible freely jointed ring this divergence turns into a
  second-order transition from a crumpled state, where the mean area
  scales as $\A\sim N$, to a smooth state with $\A\sim N^2$.  In the
  current work we extend these two models to the case where the
  swelling of the ring is caused by trapped ideal-gas particles.  The
  Gaussian model is solved exactly, and the freely jointed one is
  treated using a Flory argument, mean-field theory, and Monte Carlo
  simulations. For a fixed number $Q$ of trapped particles the
  criticality disappears in both models through an unusual mechanism,
  arising from the absence of an area constraint. In the Gaussian case
  the ring swells to such a mean area, $\A\sim NQ$, that the pressure
  exerted by the particles is at $\Pc$ for any $Q$. In the freely
  jointed model the mean area is such that the particle pressure is
  always higher than $\Pc$, and $\A$ consequently follows a single
  scaling law, $\A\sim N^2f(Q/N)$, for any $Q$. By contrast, when the
  particles are in contact with a reservoir of fixed chemical
  potential, the criticality is retained. Thus, the two ensembles are
  manifestly inequivalent in these systems.
  }

\maketitle

\section{Introduction}  \label{secIntroduction}

Polymer rings confined to two dimensions (2D) have been the subject of
several theoretical works
\cite{Butler1987,Wiegel1988,Duplantier1989,Duplantier1990,Cardy1994,Fisher1987,Fisher1990,Fisher1990_2,Rudnick1991,Rudnick1993,Levinson1992,Maritan1993,Gaspari1993,Haleva2006},
in part as highly idealized models for membrane vesicles. The
statistics of 2D chains have been studied also experimentally, using
polymers adsorbed at liquid interfaces
\cite{Vilanove1980,Gavranovic2005} or membranes \cite{Maier2000}, as
well as vibrated granular chains \cite{BenNaim2001} and rings
\cite{BenNaim2006}. In those experimental systems both random-walk and
self-avoiding-walk statistics were observed.

Pressurized 2D rings were theoretically investigated in a number of
works
\cite{Fisher1987,Fisher1990,Fisher1990_2,Rudnick1991,Rudnick1993,Levinson1992,Maritan1993,Gaspari1993,Haleva2006}.
In the Rudnick-Gaspari model
\cite{Rudnick1991,Rudnick1993,Levinson1992} the ring is represented by
a closed Gaussian chain of $N$ springs subject to an inflating
pressure differential $p$. This model yields a divergent mean area at
a critical pressure, $\Pc\sim N^{-1}$.  Recently we have demonstrated
that the inclusion of chain inextensibility (\ie considering a freely
jointed chain rather than a Gaussian one) turns this divergence into a
second-order smoothening transition \cite{Haleva2006}. For $p<\Pc$ the
ring is in a crumpled, random-walk state with the mean area scaling as
$\A\sim Nf^<(pN)$, whereas for $p>\Pc$ the ring is smooth with $\A\sim
N^2f^>(pN)$.

In many practical cases a pressure difference across a closed surface
is achieved osmotically through a concentration difference, \eg by
trapping particles inside the envelope.  In the current study we
extend the models of pressurized Gaussian and freely jointed 2D rings
to the case where the swelling is caused by a trapped ideal gas of
particles. We examine both the canonical and grand-canonical
ensembles, in which the particle number $Q$ or the particle fugacity
$\gamma$ are fixed, respectively.

The model is defined in the following section. In Section \ref{secGC}
we derive exact results for a Gaussian ring swollen by trapped
particles. We then treat the swelling of a freely jointed ring in
Section \ref{secFJC} by employing a Flory argument, mean-field
calculation and Monte Carlo simulations. In addition, we derive exact
asymptotes for the large-swelling regime.  Finally, the results are
discussed in Section \ref{secSummary}.

\section{Model}\label{secModel}

We model a 2D polymer ring as a closed ideal chain of $N$ monomers and
no bending rigidity. The ring encapsulates an ideal gas of particles.
We shall consider separately the two cases of a Gaussian and a freely
jointed chain. The unperturbed monomer--monomer link length is taken
as the unit length, $l\equiv 1$. For a freely jointed chain this
length remains fixed under swelling. For a Gaussian chain, however,
the link length has a statistical distribution that varies with
swelling; $l$ is then defined as the unperturbed root-mean-square link
length.

In a canonical ensemble, where the number $Q$ of particles is fixed,
the partition function is given, up to a constant prefactor, by
\begin{equation}
  Z(N,Q) = \int_0^\infty  dA P_0(N,A) A^Q/Q!,
\label{eqZNQ}
\end{equation}
where $A$ is the area bounded by the ring, and $P_0(N,A)$ is the
probability distribution function of the area for an unpressurized
$N$-monomer ring (for either a Gaussian or a freely jointed
chain). The mean area of the ring is given by
\begin{equation}
  \label{eqANQ}
  \langle A(N,Q)\rangle = (Q+1) Z(N,Q+1)/Z(N,Q).
\end{equation}

In a grand-canonical ensemble, where the particles are in contact with
a reservoir of fixed fugacity $\gamma$, we use \Eq{eqZNQ} to write
the grand partition function as
\begin{equation}
  \Z(N,\gamma) = \sum_{Q=0}^{\infty}Z(N,Q) \gamma^Q = 
  \int_0^\infty dA P_0(N,A) e^{\gamma A}.
  \label{eqZGrC}
\end{equation}
The mean area in this ensemble is given, therefore, by
\begin{equation}
  \langle A(N,\gamma)\rangle = \partial{\ln\Z}/\partial{\gamma},
  \label{eqAGrC}
\end{equation}
and the mean particle number by
\begin{equation}
  \langle Q(N,\gamma)\rangle = \partial{\ln\Z}/\partial{\ln\gamma}=
  \gamma\A.
  \label{eqQGrC}
\end{equation}
Thus, the mean particle density, $c=\langle Q\rangle/\A$, is equal to
the fugacity $\gamma$, as it should for an ideal gas.

The correlation between fluctuations in particle number and in the
ring area is characterized by the following covariance,
\begin{equation}
  C_{QA}(N,\gamma)=\frac{\langle QA\rangle -\langle Q\rangle\A}{\langle Q\rangle\A}.
\end{equation}
From \Eq{eqZGrC} we find
\begin{equation}
  \langle Q A\rangle=
  \frac{1}{\Z}\frac{\partial^2\Z}{\partial(\ln\gamma)\partial\gamma}=
  \gamma\langle A^2\rangle,
  \label{eqQAGrC}
\end{equation}
which, combined with \Eq{eqQGrC}, yields
\begin{equation}
  C_{QA}(N,\gamma)= \frac{\langle\Delta A^2\rangle}{\A^2},
  \label{eqCorrelateGrC}
\end{equation}
where $\langle\Delta A^2\rangle = \langle A^2\rangle - \A^2$ is the
mean-square area fluctuation.  Hence, interestingly, the
cross-correlation of $Q$ and $A$ is identical to the relative
mean-square fluctuation in the ring area.

Our model system consists of two subsystems --- particles and ring
monomers --- which may have different relaxation times. We note,
however, that in \Eqs{eqZNQ}{eqZGrC} all configurations of both
subsystems are traced over. Hence, the equilibrium properties studied
in this paper are unaffected by such time-scale differences.

Several complications related to the definition of the ring area in
the model should be mentioned. In principle the $A$ appearing in
\Eq{eqZNQ} should be the {\em geometrical} area $A_{\rm g}$ of the
ring, since it is $A_{\rm g}$ that determines the translational
entropy of the particles. As in previous works
\cite{Rudnick1991,Rudnick1993,Levinson1992,Haleva2006}, however, we
are technically bound to use the {\em algebraic} area $A$ instead, \ie
the area obtained from an integral over the ring contour.  This area
may contain both positive and negative contributions, $A_+$ and $A_-$.
In an unperturbed ring positive and negative areas are equally
favorable, and the mean algebraic area then vanishes. (The algebraic
area may also ``count'' a certain geometrical-area contribution more
than once due to chain winding.) On the one hand, particle entropy
forces us to use only the positive-area part of $P_0$ (hence the
integration from $0$ to $\infty$ in \Eq{eqZNQ}). Consequently, the
mean algebraic area of an unperturbed ring, $\A$, does not vanish.
This is the definition used in the analytical parts of this work. In
the simulations, on the other hand, we follow changes in the actual
algebraic area of the ring, allowing the total area to become
negative. Thus, in the numerical parts of this work $\A$ of an
unperturbed ring does vanish.  In addition, the simulated particles
are placed only inside positive parts of the ring area, \ie their
entropy-relevant area is $A_+\neq A$. Finally, in \Eq{eqZGrC} one
notices a direct analogy between the problem of fixed particle
fugacity studied here and that of an empty ring subject to a fixed
pressure, which was studied in \cite{Rudnick1991,Haleva2006}. The
mapping is not exact, however, since in the current case $A$ is
restricted to positive values, whereas in
\cite{Rudnick1991,Haleva2006} it was not. All these subtleties are
significant only in the weak-swelling regime of small areas. As the
inner pressure exerted by the particles increases, the distinction
between the various area definitions becomes negligible and does not 
affect our main results as will be demonstrated below, as will be 
demonstrated below.

\section{Gaussian Ring} \label{secGC}

In this section we consider a 2D Gaussian ring swollen by trapped
particles, for which exact results can be derived. The chain consists
of a set of $N$ springs with fixed spring constant, $\lambda=1$, in
units of $\KT/l^2$ (to yield an unperturbed root-mean-square spring
length of $l=1$). For the sake of the following sections, in which the
spring constant is allowed to change, we keep the results dependent on
$\lambda$ without substituting $\lambda =1$.

\subsection{Canonical Ensemble}

The probability distribution function of the algebraic area for a 2D
Gaussian ring was calculated by Khandekar and Wiegel
\cite{Wiegel1988,Duplantier1989},\footnote{Equation (\ref{eqPofA})
differs from the original Khandekar-Wiegel expression by a factor of
$2$, since the particle translational entropy in the current model
restricts our analysis to positive areas (Eq.\ (\ref{eqZNQ})).}
\begin{equation}
  \label{eqPofA}
  P_0^{\rm G}(N,A)=\frac{1}{\lambda^N}\frac{2\pi\lambda}{N \cosh^2
  (2\pi\lambda A/N)}.
\end{equation}
Substituting \Eq{eqPofA} in \Eq{eqZNQ}, we obtain the exact
partition function,
\begin{equation}
  \label{eqZGC}
    Z^{\rm G}(N,Q) = \frac{2}{\lambda^N} \left(\frac{N}{4\pi\lambda}\right)^Q(1-2^{1-Q})\zeta(Q),
\end{equation}
where $\zeta$ is the Riemann zeta function. The mean area is then
given, according to \Eq{eqANQ}, by
\begin{equation}
  \langle A^{\rm G}(N,Q)\rangle =\frac{N(Q+1)}{4\pi\lambda}\frac{1-2^{-Q}}{1-2^{1-Q}}\frac{\zeta(Q+1)}{\zeta(Q)},
  \label{eqAGC}
\end{equation}
which turns, in the limit of large $Q$, to 
\begin{equation}
  \label{eqAGCRG}
  \langle A^{\rm G}(N,Q\gg1)\rangle =  N Q / (4\pi \lambda).
\end{equation}

In the current case of fixed $\lambda$, we thus get that the ring
swells gradually (linearly) with $Q$. This is qualitatively different
from the Rudnick-Gaspari result for the swelling as a function of
fixed pressure \cite{Rudnick1991,Rudnick1993,Levinson1992}, where the
mean area diverges at a finite critical pressure, $\Pc=4\pi/N$ (in
units of $\KT/l^2$). The ability of the Gaussian ring to swell
indefinitely stems, as in the Rudnick-Gaspari model, from its
extensibility.

Since the mean area is proportional to $Q$, we obtain in this ensemble
the peculiar result that the particle density is independent of
particle number,
\begin{equation}
  c^{\rm G}(N) = Q/\langle A^{\rm G}\rangle = 4\pi\lambda/N.
  \label{eqCGC}
\end{equation}
The same holds for the particle fugacity,
\begin{equation}
  \gamma^{\rm G}(N) = \exp(-\partial\ln Z^{\rm G}/\partial Q) = 4\pi \lambda/N,
  \label{eqFuGC}
\end{equation}
which is equal to the density of \Eq{eqCGC}, as expected for an ideal
gas.  The mean pressure of the gas, in our dimensionless units, is
$p=c=4\pi/N$.  (Recall that $\lambda=1$.)  Thus, for any $Q$ the ring
swells to such an extent that the pressure exerted on it by the gas is
always at the Rudnick-Gaspari $p_{\rm c}$, yet without the associated
criticality.

\subsection{Grand-Canonical Ensemble}

The grand partition function is obtained from \Eqs{eqZGrC}{eqPofA} as
\begin{equation}
  \Z^{\rm G}(N,\gamma) = \frac{1}{\lambda^{N+1}} 
  \left[-\lambda + \psi(-x) - \psi(1/2-x) \right],
  \label{eqZGCGC}
\end{equation}
where $x\equiv\gamma N/(8\pi\lambda)$, and $\psi$ is the digamma
function (the logarithmic derivative of the gamma function). The mean
area is calculated according to \Eq{eqAGrC}, yielding
\begin{equation}
  \begin{split}
  \langle A^{\rm G}(N,\gamma)\rangle =& \frac{1}{\lambda^N\Z^{\rm G}}
  \frac{N}{4\pi\lambda}
  \left\{ \psi(-x) - \psi(1/2-x) \right.\\
  &\left. - x \left[\psi'(-x) -\psi'(1/2-x) \right]\right\},
  \end{split}
  \label{eqAofMu}
\end{equation}
where prime denotes a first derivative.  The mean number of trapped
particles, according to \Eq{eqQGrC}, is simply given by $\langle
Q^{\rm G}\rangle=\gamma\langle A^{\rm G}\rangle$.

The digamma function and its derivative diverge for small arguments as
$\psi(x)\simeq -x^{-1}$ and $\psi'(x)\simeq x^{-2}$, respectively.
Thus, the expressions for $\Z^{\rm G}$, $\langle A^{\rm G}\rangle$,
and $\langle Q^{\rm G}\rangle$ diverge at a critical fugacity (with
$\lambda = 1$),
\begin{equation}
  \Gc = 4\pi/N.
\end{equation}
Since for the ideal gas $p=c=\gamma$, this divergence is analogous to
that of the Rudnick-Gaspari model at $\Pc=4\pi/N$.

\section{Freely Jointed Ring} \label{secFJC}

We now turn to the case of a freely jointed chain, \ie an inextensible
ring whose link lengths are fixed.

\subsection{Canonical Ensemble}
\subsubsection{Flory Argument} \label{secFA}

We begin by examining a simple Flory argument, following the lines of
\cite{Haleva2006}. The free energy of the ring (in units of $\KT$) is
expressed as a function of $R$, the radius of the statistical cloud of
monomers (\ie the mean radius of gyration).  We divide it into three
terms,
\begin{eqnarray}
  \label{eqFloryFJC}
  &&F(R,Q) = F_{\rm el} + F_{\rm inext} + F_{\rm en}, \\ 
  &&F_{\rm el} \sim R^2/N,\ \ 
  F_{\rm inext}  \sim R^4/N^3,\ \ 
  F_{\rm en} \sim Q [\ln (Q/R^2)-1]. \nonumber
\end{eqnarray}
The first term is the usual entropic-spring free energy of a Gaussian
chain \cite{PolymerPhysics2003}. The second is the leading
non-Gaussian correction due to inextensibility (see App.\ A of
\cite{Haleva2006}). The last term comes from the translational entropy
of the ideal gas, where the mean area of the ring is taken
proportional to $R^2$ \cite{Duplantier1990,Cardy1994}.

Unlike the case of fixed pressure \cite{Haleva2006}, the free energy
of \Eq{eqFloryFJC} exhibits no phase transition with increasing
$Q$. Upon minimization with respect to $R$ we get the following
scaling law:
\begin{equation}
  R^2 \sim N^2 f(Q/N),
\label{eqFR2}
\end{equation}
which, despite the crudeness of the Flory argument, turns
out to be correct, as demonstrated below.

\subsubsection{Mean-Field Approximation} \label{secMF}

We proceed to study the freely jointed model by relaxing the strict
constraints of fixed link lengths into harmonic potentials. This kind
of approximation, used in several previous polymer theories
\cite{Harris1966,Thirumalai1995,Diamant2000,Haleva2006}, is equivalent
to a mean-field assumption \cite{Thirumalai1995}.  The calculation is
identical to that for a Gaussian chain (Sec.\ \ref{secGC}), yet the
spring constant $\lambda$ is not fixed but determined
self-consistently to maintain the root-mean-square link length equal
to $l=1$.

To impose the relaxed inextensibility constraint we differentiate
$Z^{\rm G}$ with respect to $\lambda$ to get the mean-square link
length, and then set it to 1,
\begin{equation}
  \label{eqUMF}
  -N^{-1} \partial\ln Z^{\rm G}/\partial\lambda = (1+\Q)/\lambda = 1,
\end{equation}
where $\Q\equiv Q/N$ is the rescaled particle number.  For
$\Q\rightarrow 0$ we get $\lambda=1$, as expected, but as the number
of trapped particles increases, the springs get stiffer according to
\begin{equation}
  \lambda=1 + \Q. 
  \label{eqL}
\end{equation}
This value is substituted in \Eq{eqAGCRG} to obtain
\begin{equation}
  \label{eqAMF}
  \langle A^{\rm FJ}(N,\Q)\rangle = \Amax \Q/(\Q+1).
\end{equation}
As $\Q\rightarrow\infty$ the mean area appropriately tends to its
maximum value, $\Amax = N^2/(4\pi)$, which is the area of a circle
with perimeter $N$.  Equation (\ref{eqAMF}) shows that the swelling of
the ring with increasing $Q$ is gradual, without any phase transition.

The particle density is given by
\begin{equation}
  \label{eqcMF}
  c^{\rm FJ}(N,\Q) = Q/\langle A^{\rm FJ}\rangle = 4\pi (\Q+1)/N.
\end{equation}
The fugacity is found from \Eqs{eqFuGC}{eqL} as
\begin{equation}
  \label{eqFuMF}
  \gamma^{\rm FJ}(N,\Q) = 4\pi (\Q+1)/N,
\end{equation}
which is equal to $c^{\rm FJ}$, as it should for an ideal gas. The
exerted pressure is $p=\gamma=(4\pi/N)(\Q+1)$, which is larger
than $\Pc = 4\pi/N$ for any $Q$.

\subsubsection{Exact Asymptote for Strong Swelling}
\label{secEx}

In the limit of large particle number, $\Q\gg 1$, the partition
function can be calculated exactly. In this limit the chain is highly
swollen and its statistics is governed by large-area conformations.
Thus, to apply \Eq{eqZNQ}, we merely need to know how the area
probability distribution function for a 2D freely jointed ring,
$P_0^{\rm FJ}(N,A)$, decays to zero as $A$ approaches $\Amax$. This
calculation is presented in Appendix A. The result is
\begin{equation}
  P_0^{\rm FJ}(N, A \lesssim \Amax) \sim (\Amax-A)^{N/2}.
  \label{eqPNAFT}
\end{equation}
Substituting \Eq{eqPNAFT} in \Eq{eqZNQ}, we get
\begin{equation}
  Z^{\rm FJ}(N,Q\gg N) = (\Amax)^{Q+N/2} \frac{\Gamma(1+N/2)}{\Gamma(1+N/2+Q)},
\end{equation}
where $\Gamma$ is the gamma function. The mean area is then obtained
from \Eq{eqANQ} as
\begin{equation}
  \langle A^{\rm FJ}(N, \Q\gg 1) \rangle \simeq \Amax\frac{\Q}{\Q+1/2}
  \simeq \Amax \left(1 - \frac{1}{2\Q} \right).
  \label{eqANq}
\end{equation}
Thus, the exact approach of the mean area to its maximum value differs
from the mean-field result, \Eq{eqAMF}, by a factor of $1/2$.

\subsubsection{Monte Carlo Simulations}\label{secMC}

\begin{figure}[tbh]
  \centering
  \vspace{1.1in}
  \includegraphics[width=2.5in]{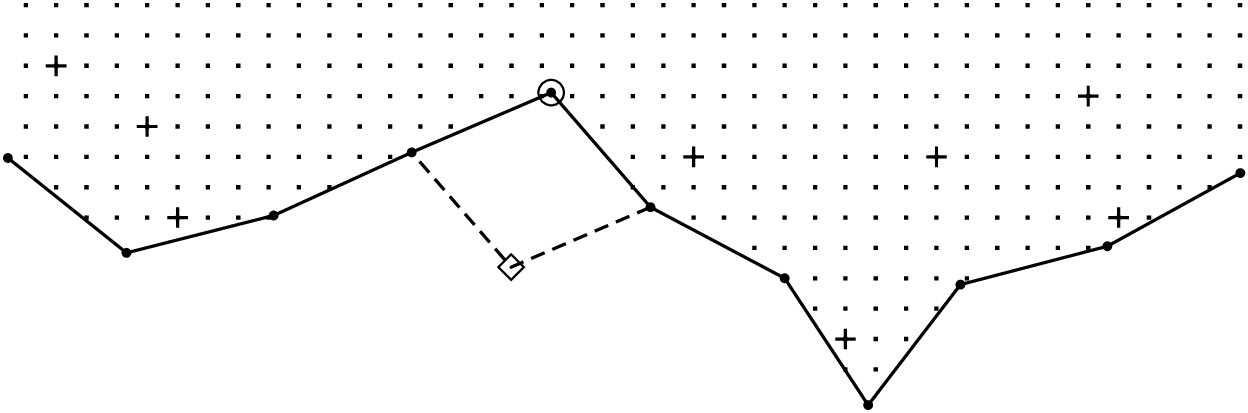}
  \vspace{0.2in}
  \caption{Schematic view of a section of the simulated system. Point
    particles (marked by $+$) are located on lattice sites confined
    inside the ring. On each step each particle is moved to a randomly
    chosen neighboring site. In the second part of the step a single
    randomly chosen chain vertex (marked by a circle) is moved to the
    only other position (marked by a diamond) that maintains the
    lengths of the two links attached to it.}
  \label{figPolygonSchema}
\end{figure}

We performed Monte Carlo simulations to obtain the mean area $\A$ of a
freely jointed ring as a function of the number $Q$ of trapped
ideal-gas particles for various ring sizes $N$. The algorithm combines
an off-lattice scheme for the polymer, as was used in
\cite{Haleva2006}, with a lattice model for the particles. The
simulated system is schematically shown in \Fig{figPolygonSchema}. The
ring is represented by a polygon of $N$ equal edges of length $l=1$.
The 2D coordinates of the vertices take continuous values. The
coordinates of the particles are defined on a 2D square lattice with
lattice constant of either $l$ or $l/5$, depending on the required
precision.

The initial configuration is a fully stretched, regular polygon of purely
positive algebraic area, $A=A_+$. During the simulation we keep track
of positive and negative contributions to $A$ and place particles only
on lattice sites belonging to $A_+$. Each step of the simulation is
composed of two actions. In the first, each particle is moved to a
randomly chosen juxtaposed lattice site, unless this site is outside
the positive-area part of the polygon. In the second action, a
randomly chosen chain vertex is moved to the only other position that
satisfies the edge-length constraint (see Fig.\
\ref{figPolygonSchema}). This action is automatically rejected if it
makes a particle leave the positive-area part of the ring. These
dynamics involve $O(Q)$ operations per step.  The number of operations
required for equilibration is $O(N^3 Q)$, limiting our investigation
to $N \lesssim 500$. The simulations were performed for $N$ between
$50$ and $400$ and $Q$ between $0$ and $4N$.

\begin{figure}[tbh]
  \centering
  \vspace{0.2in}
  \includegraphics[width=3.3in]{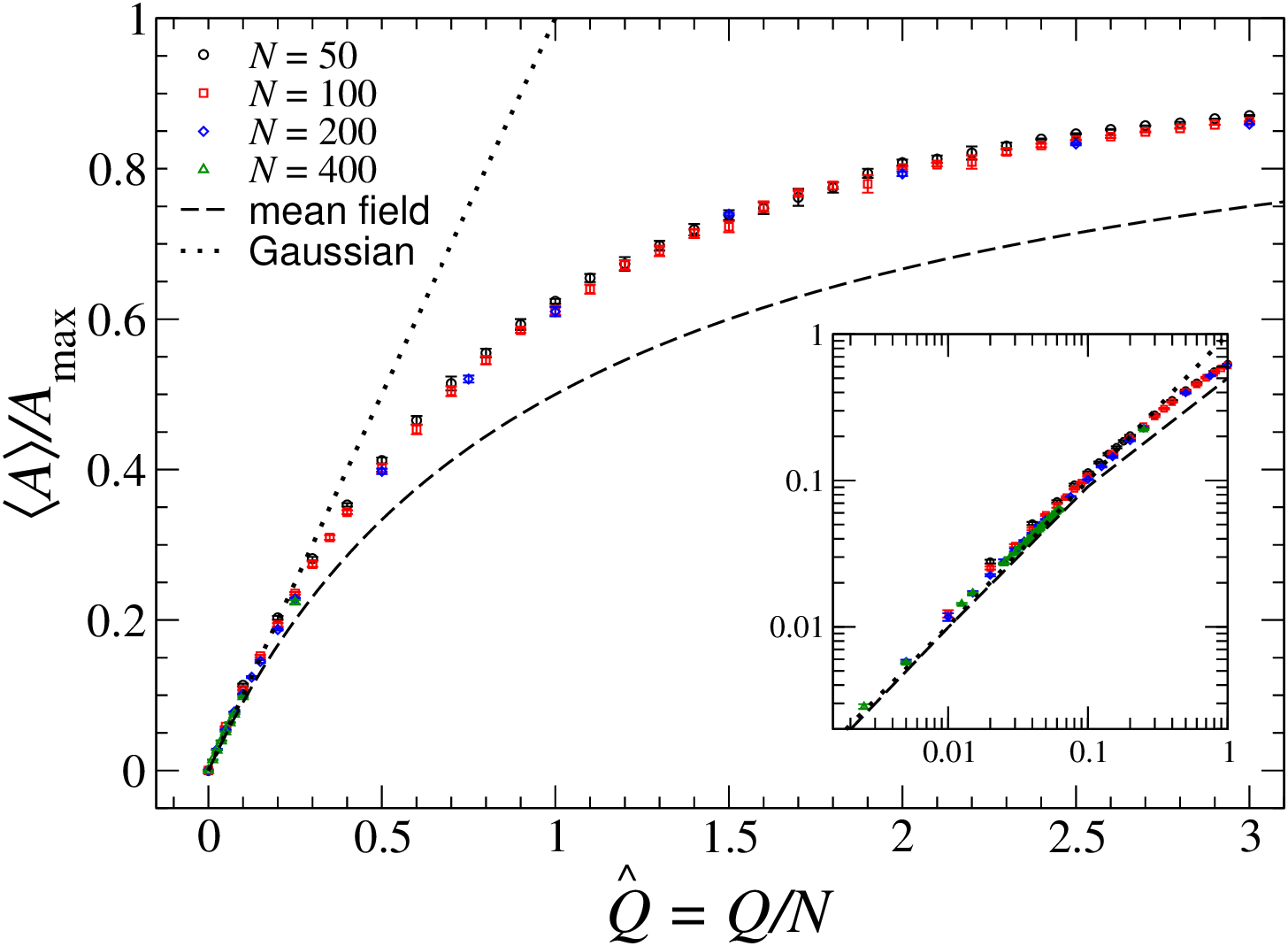}
  \caption{(Color online) Mean area of a freely jointed ring, 
    scaled by $\Amax\sim N^2$, as a function of the rescaled number of
    particles. Monte Carlo simulation results for various values of
    $N$ are represented by symbols with error bars. The dashed line
    shows the mean-field result (Eq.\ (\ref{eqAMF})). The result for a
    Gaussian ring (Eq.\ (\ref{eqAGCRG})) is given for reference (dotted
    line).  The inset focuses on the small $\Q$ region using a logarithmic
    scale.}
  \label{figAtoq}
\end{figure}

Figure \ref{figAtoq} shows the simulation results for the mean area
(scaled by $\Amax\sim N^2$) as a function of the rescaled particle
number $\Q=Q/N$. All data for different values of $N$ collapse onto a
single universal curve, in accord with the scaling law obtained from
the Flory argument and mean-field calculation (Eqs.\ (\ref{eqFR2}) and
(\ref{eqAMF})). Yet, only for small $\Q$ does this curve coincide with
the mean-field scaling function.  For such small $\Q$ it coincides
also with the Gaussian result, (Eq.\ (\ref{eqAGCRG}) with $\lambda
=1$).

\begin{figure}[tbh]
  \centering
  \vspace{0.3in}
  \includegraphics[width=3.3in]{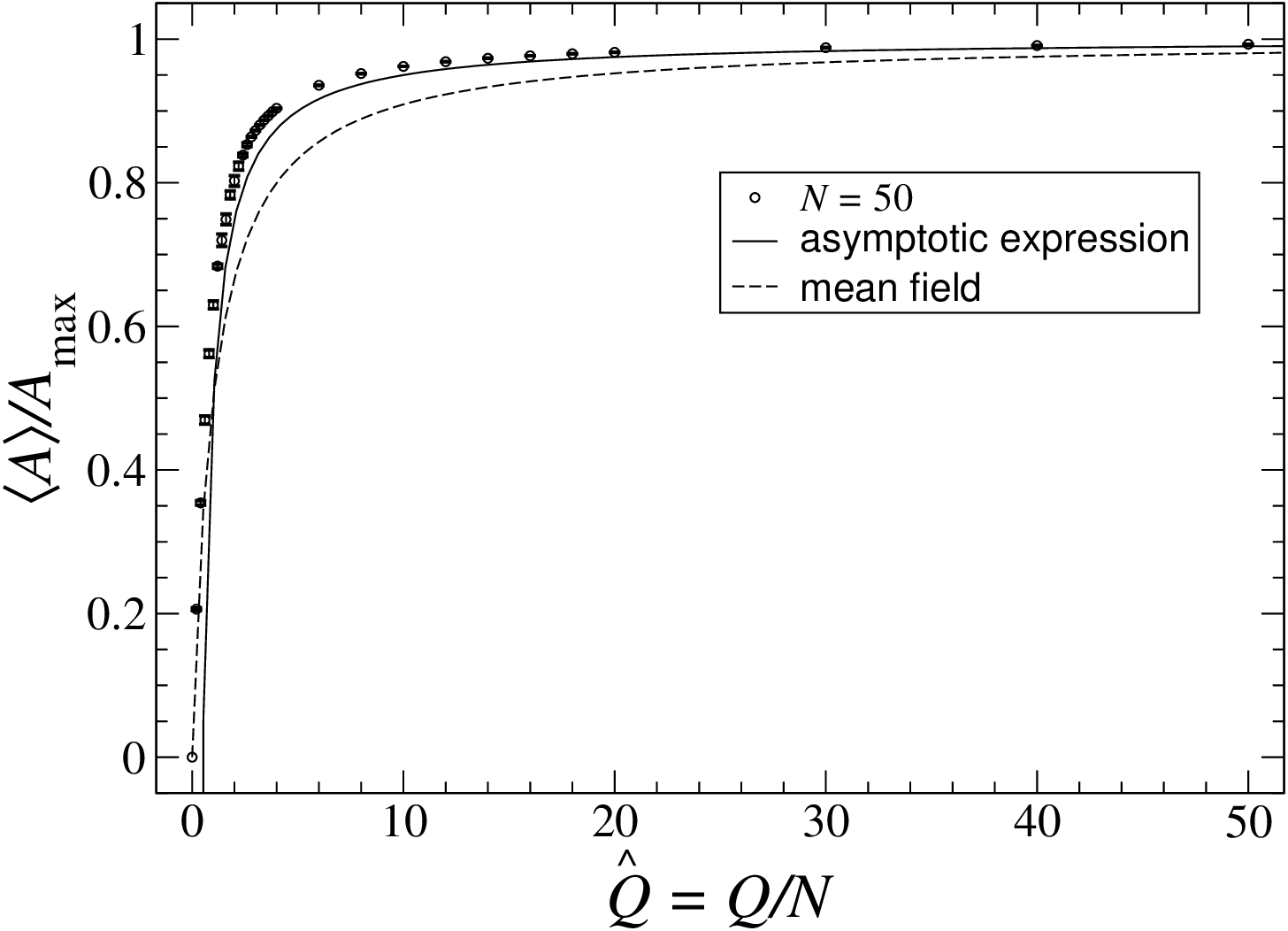}
  \caption{Mean area of a freely jointed ring, scaled by $\Amax$, 
    as a function of the rescaled particle number, as obtained by
    Monte Carlo simulations for $N=50$ (symbols with error bars). The
    solid line shows the exact asymptote for a highly swollen ring
    (Eq.  (\ref{eqANq})). The dashed line presents the mean-field
    result (Eq.\ (\ref{eqAMF})).}
  \label{figAtoLargeq}
\end{figure}

To confirm the highly swollen behavior derived in Section \ref{secEx}
we simulated rings of $N=50$ with $Q$ between $0$ and $50N$.  The
results are presented in \Fig{figAtoLargeq}, showing good agreement
for large $\Q$ with the exact asymptote, \Eq{eqANq}.

\subsection{Grand-Canonical Ensemble}

\subsubsection{Flory Argument} \label{secGrFA}

To account for contact of the particles with a reservoir of fixed
fugacity $\gamma$ we add a $- Q\ln\gamma$ term to the Flory free
energy, \Eq{eqFloryFJC}. Minimization with respect to $Q$ yields the
following grand potential:
\begin{eqnarray}
  \label{eqFloryGrcGC} 
        &&{\cal F}(R,\gamma) = F_{\rm el} + F_{\rm inext} + F_{\rm res},\\ 
        &&F_{\rm el} \sim R^2/N,\ \ 
        F_{\rm inext} \sim R^4/N^3,\ \ 
        F_{\rm res} \sim -\gamma R^2. \nonumber
\end{eqnarray} 
This Landau-like free energy has a second-order phase transition at
$\Gc\sim N^{-1}$, similar to the one obtained for fixed pressure in
\cite{Haleva2006}. For $\gamma<\Gc$ $F_{\rm inext}$ is negligible, and
$R$ has a Gaussian distribution with $\langle R^2\rangle \sim
N(1-\gamma/\Gc)^{-1}$. For $\gamma>\Gc$ we have $R^2\sim
N^2(\gamma/\Gc - 1)$.

\subsubsection{Mean-Field Approximation}

As in \Sec{secMF} we apply a relaxed inextensibility constraint by
determining the spring constant $\lambda$ so that the root-mean-square
link length should be $l=1$. In the current grand-canonical case this
is done by differentiating $\Z^{\rm G}$ of \Eq{eqZGCGC} with respect
to $\lambda$,
\begin{equation}
  \label{eqUGrC}
  \begin{split}
    -&N^{-1} \partial\ln \Z^{\rm G}/\partial\lambda  \\
    =& \frac{1}{\lambda} + \frac{2}{\lambda^{N+1}N\Z^{\rm G}} x
    \left\{ \psi(-x) - \psi(1/2-x) \right.\\
    &\left. - x \left[\psi'(-x) - \psi'(1/2-x) \right] \right\}
    =1,
  \end{split}
\end{equation}
thus obtaining a transcendental equation for $\lambda(\gamma,N)$.
Equation (\ref{eqUGrC}) can be combined with \Eq{eqAofMu} to get a
simpler expression for $\A$ as a function of $\lambda$,
\begin{equation}
  \label{eqMeanArea2}
  \langle A^{\rm FJ}(N,\gamma,\lambda)\rangle = N(\lambda - 1)/\gamma.
\end{equation}
Numerical solution of \Eq{eqUGrC} for $\lambda$ and substitution of
the result in \Eq{eqMeanArea2} yield the mean area as a function of
$\gamma$ and $N$.

We can get a good approximation for $\lambda(\gamma,N)$ by
substituting for the diverging terms in \Eq{eqUGrC} $\psi(x)\simeq
-x^{-1}$ and $\psi'(x)\simeq x^{-2}$.  This gives
\begin{equation}
  \label{eqLambda}
  \lambda(\hat\gamma,N\gg 1) \simeq \frac{\hat\gamma+1+\frac{1}{N} 
  + \sqrt{(\hat\gamma-1)^2 +\frac{2}{N}(\hat\gamma+1)+\frac{1}{N^2}}}{2},
\end{equation}
where $\hat\gamma\equiv\gamma N/(4\pi)$ is the rescaled fugacity.
This result for $\lambda$ is the same as the one obtained for fixed
pressure $p$ \cite{Haleva2006} with $p=\gamma$. In the thermodynamic
limit, defined here as $N\rightarrow\infty$ and $\gamma\rightarrow 0$
such that $\hat\gamma$ is finite, \Eq{eqLambda} reduces to the
continuous but nonanalytic function,
\begin{equation}
  \label{eqLambdaLimit}
  \lambda(\hat\gamma,N\rightarrow\infty) = 
  \begin{cases} 
    1 & \hat\gamma<1\\ 
    \hat\gamma & \hat\gamma>1.
  \end{cases}
\end{equation}

Substituting \Eq{eqLambda} in \Eq{eqMeanArea2} yields an approximate
expression for $\A$ as a function of $\hat\gamma$ and $N$,
\begin{equation}
  \label{eqMeanArea3}
  \begin{split}
    &\langle A^{\rm FJ}(\hat\gamma\not\ll 1,N\gg 1)\rangle \simeq \\
    &\frac{N^2}{4\pi}\frac{\hat\gamma-1+\frac{1}{N} + \sqrt{(\hat\gamma-1)^2
    +\frac{2}{N}(\hat\gamma+1)+\frac{1}{N^2}}}{2\hat\gamma},
  \end{split}
\end{equation}
which in the thermodynamic limit becomes
\begin{equation}
  \label{eqAGrCFJCLim}
  \langle A^{\rm FJ}\rangle =
  \begin{cases} 
    \frac{N}{4\pi}\frac{1}{\hat\gamma(1-\hat\gamma)}\xrightarrow{\hat\gamma\rightarrow 1^-} 
    \frac{N}{4\pi}\frac{1}{1-\hat\gamma} & 1-\hat\gamma \gg N^{-1/2}\\
    \frac{N^{3/2}}{4\pi} & |\hat\gamma-1| \ll N^{-1/2}\\
    \frac{N^2}{4\pi}\frac{\hat\gamma-1}{\hat\gamma} 
    \xrightarrow{\hat\gamma\rightarrow 1^+}\frac{N^2}{4\pi}(\hat\gamma-1)  & \hat\gamma -1 \gg N^{-1/2},
  \end{cases}
\end{equation}
thus exhibiting a continuous phase transition at $\hat\gamma_{\rm
c}=1$, analogous to the one found for fixed pressure (Eq. (16) in
\cite{Haleva2006}). As shown in \Eq{eqQGrC}, the mean number of
trapped particles is given by $\langle Q^{\rm FJ}\rangle =
\gamma\langle A^{\rm FJ}\rangle$ and, therefore, undergoes the same
transition at $\hat\gamma_{\rm c}$.

\subsubsection{Exact Asymptote for Strong Swelling}

As demonstrated above, and also confirmed by simulations in the next
section, fixing the fugacity in the grand-canonical ensemble is
analogous to fixing the pressure in an empty ring. Thus, to get the
asymptotic swelling at high fugacity we can readily use the exact
asymptote for high pressure derived in \cite{Haleva2006} and replace
$p\rightarrow\gamma$. This yields
\begin{equation}
\label{eqI0}
  \langle A(\hat\gamma\gg 1,N)\rangle \simeq \Amax \frac{I_1(\hat\gamma)}
  {I_0(\hat\gamma)} \simeq \Amax \left(1 - \frac{1}{2\hat\gamma} \right),
\end{equation}
where $I_n$ are modified Bessel functions of the first kind.

\subsubsection{Monte Carlo Simulations} \label{secGrMC}

We performed grand-canonical Monte Carlo simulations to check the
results of the preceding sections. To each step of the algorithm
described in \Sec{secMC} a third action is added as follows. Either a
randomly chosen particle is removed with probability $Q/(\gamma A_+)$,
or a particle is added to a randomly chosen site within the
positive-area part of the ring with probability $\gamma A_+/(Q+1)$.

\begin{figure}[tbh]
  \centering \vspace{0.4cm} { \label{figCollapseUnder}
  \includegraphics[width=3.2in]{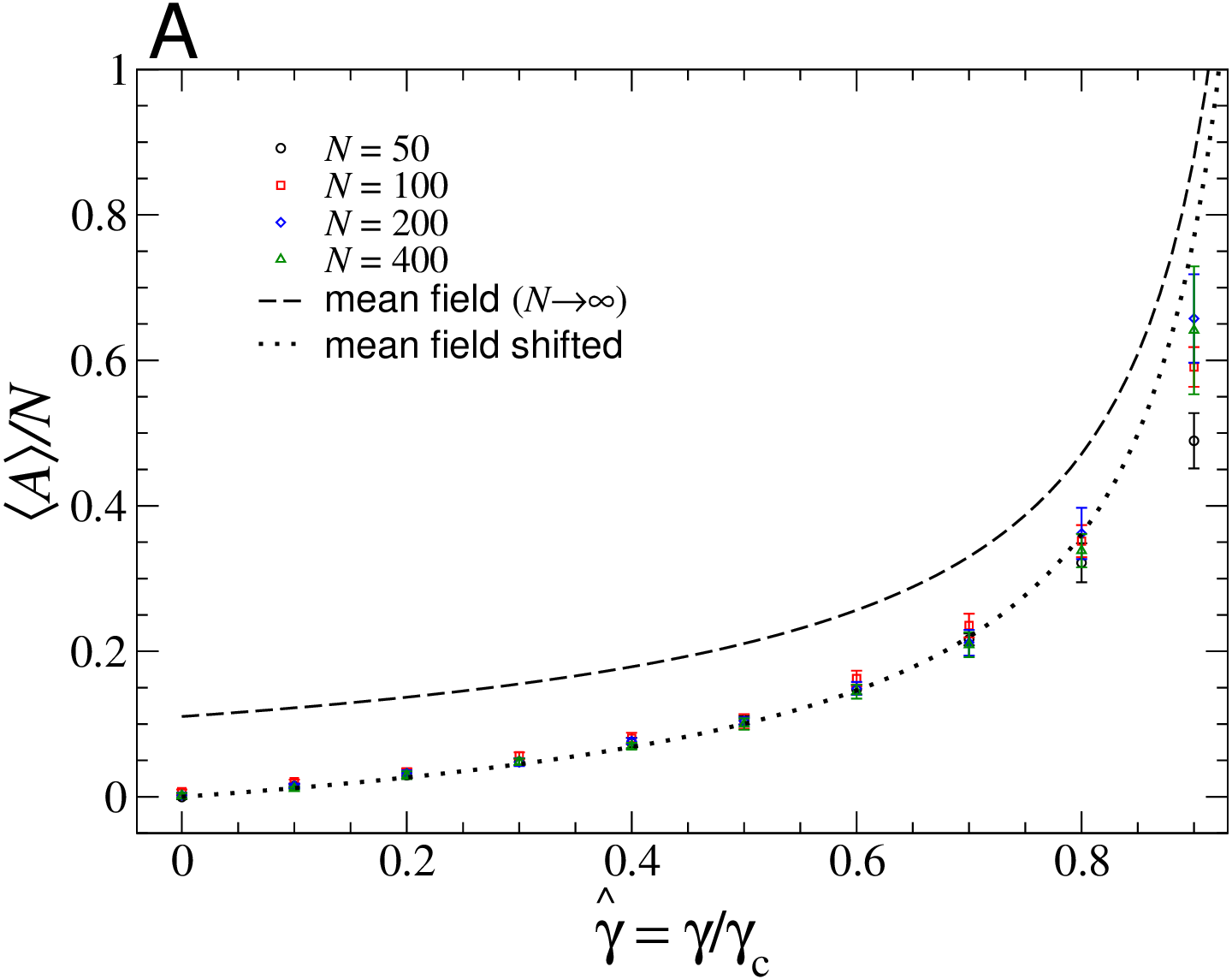} } \vspace{0.6cm} {
  \label{figCollapseOver} \includegraphics[width=3.2in]{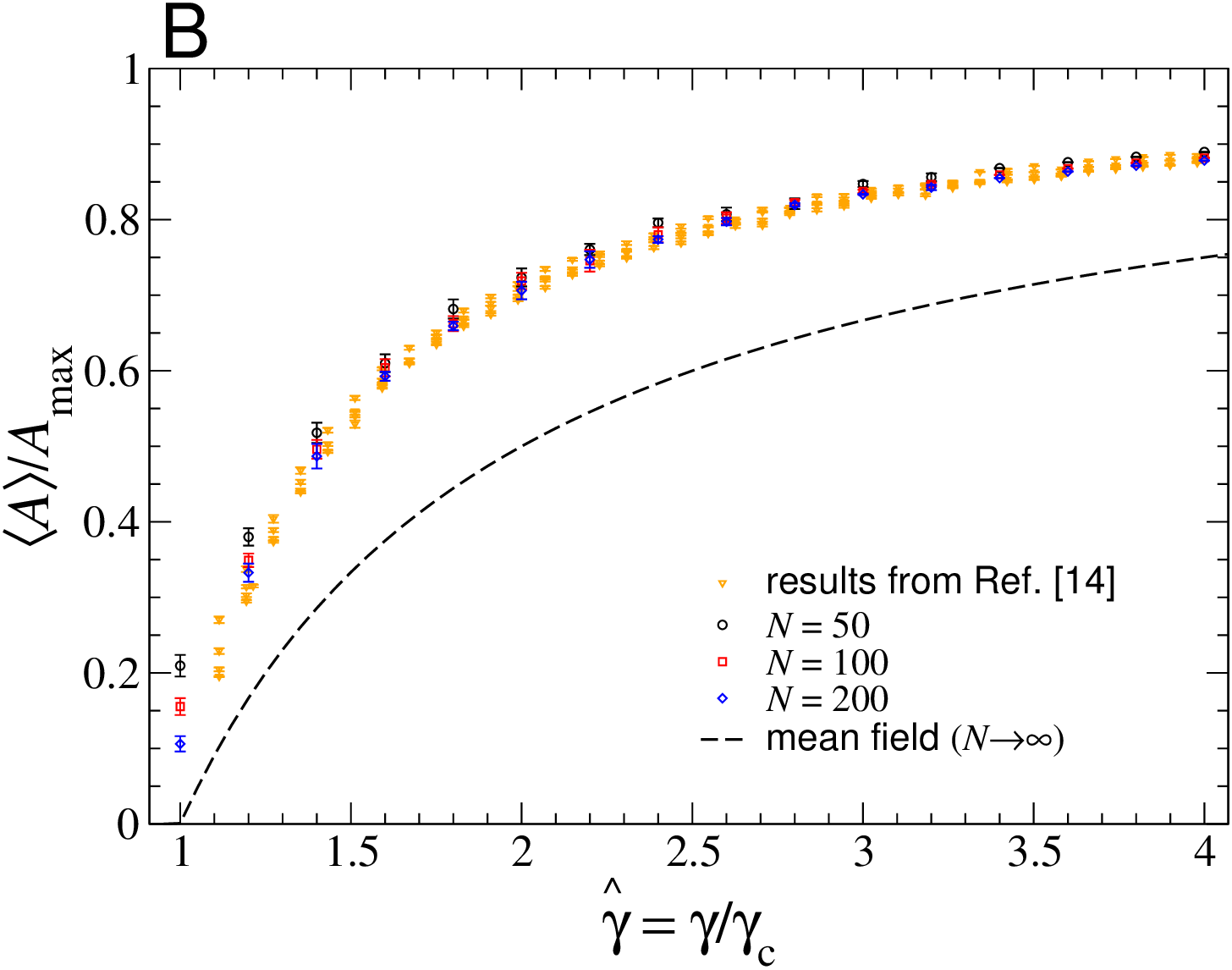} }
  \caption{(Color online) Mean area of a freely jointed ring as a
  function of particle fugacity below (A) and above (B) the critical
  point. The fugacity is scaled by $\Gc=4\pi/N$, and the area by $N$
  in (A) and by $A_{\rm max}=N^2/(4\pi)$ in (B). Symbols with error
  bars show the results of grand-canonical Monte Carlo simulations for
  different values of ring sizes $N$. The dashed lines show the
  mean-field prediction in the limit $N\rightarrow\infty$. The dotted
  line in (A) presents the mean-field result shifted down by $\langle
  A(0)\rangle/N=(\ln 2)/(2\pi)$. Panel (B) shows also simulation
  results from \cite{Haleva2006} for the swelling of a freely jointed
  ring with pressure $p$, for which the horizontal axis represents
  $\hat{p}=p/\Pc$.}  \label{figCollapseOfData}
\end{figure}

\Fig{figCollapseOfData} shows simulation results for different values
of $N$ in the crumpled and smooth states, $\hat\gamma <1$ and
$\hat\gamma>1$, respectively. The data collapse onto a single
universal curve in each state. In the weak-swelling state (Fig.
\ref{figCollapseOfData}A) the mean area scales as $\A=
Nf^<(\hat\gamma)$, whereas in the strong-swelling state (Fig.
\ref{figCollapseOfData}B) it scales as $\A=N^2f^>(\hat\gamma)$. These
results confirm the scaling laws predicted by the Flory argument and
mean-field theory, albeit with different scaling functions.  The
discrepancy in \Fig{figCollapseOfData}A between the simulation results
and the mean-field expression stems from the different definitions of
the ring area discussed in \Sec{secModel}. In the simulation we
measure the actual algebraic area, which vanishes at $\gamma =0$.  The
calculation, however, considers only positive algebraic areas (cf.\
Eq.\ (\ref{eqZNQ})) and, hence, yields a finite mean area at zero
fugacity, $\langle A(0)\rangle=N(\ln 2)/(2\pi)$.  When this value of
$\langle A(0)\rangle$ is subtracted from the mean-field result (dotted
line in Fig.\ \ref{figCollapseOfData}A), the agreement with the
simulations is excellent, indicating that the mean-field theory does
accurately capture the swelling for $\gamma<\Gc$.  On the
other hand, the disagreement between simulation and theory in
\Fig{figCollapseOfData}B arises from the inadequacy of the mean-field
approximation for stretched freely jointed chains, as was observed
also in \cite{Haleva2006}.  In \Fig{figCollapseOfData}B we compare the
data for a freely jointed chain subject to a pressure $p>\Pc$
\cite{Haleva2006} with those for fixed fugacity $\gamma > \Gc$. The
two data sets are practically indistinguishable once one identifies
$p$ with $\gamma$.

Due to computation limitations we have not directly confirmed the
asymptotic strong-swelling behavior at large $\hat\gamma$ as given by
\Eq{eqI0}; the equivalent asymptote for high pressure was confirmed in
\cite{Haleva2006}.

\begin{figure}[tbh]
  \centering
  \vspace{0.2in}
  \includegraphics[width=3.3in]{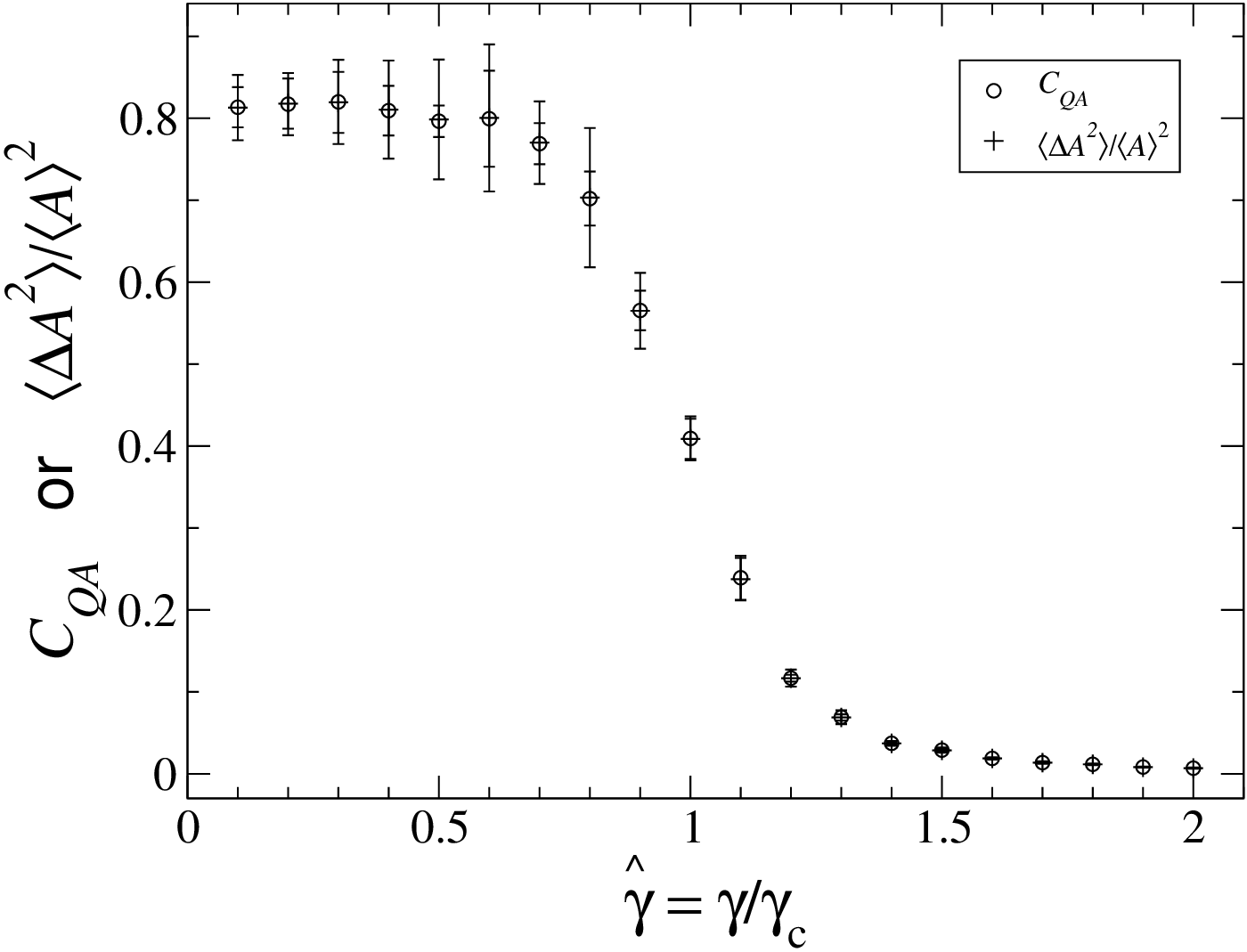}
  \caption{Correlation between particle number and ring
    area as a function of particle fugacity, compared to the relative
    mean-square area fluctuations.  Simulations were performed for an
    $N=50$ ring, and the ring area was taken as $A_+$.}
  \label{figCorrelations}
\end{figure}

To test \Eq{eqCorrelateGrC} for the cross-correlation of particle
number and ring area we measured from the simulations the covariance
$C_{QA}=(\langle Q A\rangle - \langle Q\rangle\A)/ (\langle
Q\rangle\A)$ for $N=50$ and varying $\gamma$. The results are
presented in \Fig{figCorrelations} alongside the relative mean-square
area fluctuations. The two data sets are indistinguishable, in
agreement with \Eq{eqCorrelateGrC}. (In these measurements the area
was taken as $A_+$.) As expected, the correlation is appreciable in
the crumpled, unpressurized state (small $\gamma$) and decays to zero
as the ring swells into a smooth circle (large $\gamma$).

\section{Discussion}\label{secSummary}

Fluctuating empty envelopes, such as the polymer rings studied
previously
\cite{Cardy1994,Fisher1987,Fisher1990,Fisher1990_2,Rudnick1991,Rudnick1993,Levinson1992,Maritan1993,Gaspari1993,Haleva2006},
have the special property that their thermodynamic state can be
defined by specifying the surface degrees of freedom and intensive
parameters only. For example, given the number of ring monomers $N$,
temperature $T$, and pressure $p$ of such an empty ring (actually, $N$
and the ratio $p/T$ suffice), one can completely characterize the
system, including its extensive mean area $\A$.  This situation is
somewhat similar to the one encountered in the thermodynamics of
electrodynamic modes in a cavity. Since the number of photons is
unconstrained, the system can be defined without specifying an
extensive parameter; given the surface properties of the cavity and
the radiation pressure, one can determine the cavity volume.

In the current work the rings are not empty but enclose a gas of
particles. Yet, we do not constrain their area, neither directly nor
via a conjugate pressure.  Hence, these systems still belong to the
special category just described. When the number $Q$ of trapped
particles is prescribed, we find that the critical behaviors,
previously reported for pressurized rings, disappear. The mechanism is
qualitatively different from ordinary ones, where criticality is
removed by disorder or fluctuations.  It lies here in the absence of
an area constraint, \ie in the freedom of these systems to select
their mean area so as to maximize the total entropy.  For a Gaussian
ring we have found that the mean area is $\A=NQ/(4\pi)$, implying that
the particles exert an effective pressure just equal to $\Pc=4\pi/N$
for any $Q>0$. The area never diverges, as it does for fixed $p$, but
grows gradually, indefinitely, with increasing $Q$. For a freely
jointed ring, by contrast, the mean area is such that the exerted
pressure remains above $\Pc$ for any $Q>0$. As a result, the swollen
ring always obeys the smooth-state scaling law, $\A\sim N^2f(Q/N)$.

When the particle fugacity $\gamma$ is prescribed, the criticality of
the two models is retained. In this case the system does not have the
freedom to select the pressure exerted on its boundary, since $p$ is
determined by $\gamma$ through the equation of state of the gas
($p=\gamma$ for an ideal gas). The models then become analogous to the
fixed-pressure ones, as studied in \cite{Rudnick1991,Haleva2006}.
Thus, the canonical and grand-canonical ensembles are inequivalent in
these systems. A vivid example is seen in the Gaussian model, where
the canonical mean fugacity $\gamma$ is found to be independent of $Q$
(Eq.\ (\ref{eqFuGC})), whereas the grand-canonical mean particle
number does vary with $\gamma$ (Eqs.\ (\ref{eqQGrC}) and
(\ref{eqAofMu})) and even diverges at $\Gc$. The former
result of a fugacity independent of particle number is clearly a
consequence of the unconstrained area.  We infer, therefore, that this
unusual property also underlies the ensemble inequivalence.

In the grand-canonical case we have derived, and confirmed in
simulations, an identity relating the correlation between particle
number and ring area with the area fluctuations (Eq.\
(\ref{eqCorrelateGrC}) and Fig.\ \ref{figCorrelations}). The only
assumption entering this derivation is the consideration of the
particles as an ideal gas. Thus, the same identity should hold for any
fluctuating envelope with unconstrained volume, enclosing an ideal gas
of particles.

This study has been restricted to self-intersecting rings. When a
self-avoidance term is added to the Flory grand potential of
\Sec{secGrFA}, the second-order transition disappears, as in the case
of fixed pressure \cite{Haleva2006}. We note, however, that the Flory
argument does not reproduce the correct scaling regimes for
pressurized self-avoiding rings as derived in
\cite{Fisher1987,Fisher1990,Fisher1990_2}.

The pressurized 2D envelopes studied here and in \cite{Haleva2006} are
evidently idealized systems. Possible implications for more realistic
systems of unconstrained volume, \eg 3D self-avoiding vesicles
\cite{Gompper1997} with pores or aquaporin channels, are currently
under study.

\vspace{0.1in}
\footnotesize{This work was supported by the US--Israel Binational
  Science Foundation (Grant no.\ 2002-271).}

\section*{Appendix A} 
\section*{Area probability distribution function of a highly swollen 
freely jointed ring}

\setcounter{equation}{0}
\renewcommand{\theequation}{A.\arabic{equation}} 

Unlike the probability distribution function of the area for a
Gaussian ring, $P_0^{\rm G}(N,A)$ of \Eq{eqPofA}, its counterpart for
a freely jointed ring, $P_0^{\rm FJ}(N,A)$, is not known
analytically. The asymptote of $P_0^{\rm FJ}$ as
$A\rightarrow \Amax$, nonetheless, can be calculated, which is the aim of
this Appendix.

A configuration of the ring is defined by a set of $N$ 2D vectors
specifying the monomer positions, $\{\R_j\}_{j=0,\ldots,N}$ with
$\R_0=\R_N$ to make the chain closed.  The partition function for a
fixed area $A$ is given by
\begin{equation}
  \begin{split}
    Z^{\rm FJ}(N,A)&= \int \prod_{j=1}^N d\R_j \delta(|\R_j-\R_{j-1}|-1) 
    \delta(A'[\{\R_j\}]- A) \\ 
    &= \int dp \int \prod_{j=1}^N d\R_j e^{i p (A'[\{\R_j\}]-A)} 
    \delta(|\R_j-\R_{j-1}|-1),
  \end{split}
  \label{eqZofNA}
\end{equation}
where $A'[\{\R_j\}]$ is the area of the configuration $\{\R_j\}$.

When the ring statistics is governed by highly swollen configurations,
the integration over $\{\R_j\}$ can be performed analytically using the
transfer-matrix technique \cite{Haleva2006}. This leads to
\begin{equation}
  Z^{\rm FJ}(N, A\lesssim\Amax) = \int dp e^{-i p A} 
  \left[2\pi I_0\left(\frac{i p N}{4\pi}\right)\right]^N,
  \label{eqZNA}
\end{equation}
where $I_0$ is the zeroth-order modified Bessel function of the first
kind.  The asymptotic expansion of $I_0$ for large arguments,
\begin{equation}
  I_0\left(\frac{ip N}{4\pi}\right) \simeq
  \left(\frac{2}{ipN}\right)^{1/2} e^{\frac{ip N}{4\pi}}.
\end{equation}
Substituting it in \Eq{eqZNA}, we get, up to a constant prefactor,
\begin{equation}
  Z^{\rm FJ}(N, A\lesssim\Amax) \sim \int dp e^{ip(\Amax- A)} (pN)^{-N/2},
\end{equation}
which, upon a simple change of variables (or, alternatively, a
stationary-phase approximation) readily yields
\begin{equation}
  Z^{\rm FJ}(N, A\lesssim\Amax) \sim (\Amax-A)^{N/2}.
\end{equation}

Thus, the area probability distribution function, which is
proportional to $Z^{\rm FJ}$, is given to leading order in $(\Amax-A)$
by
\begin{equation}
  P_0^{\rm FJ}(N, A \lesssim \Amax) \sim (\Amax-A)^{N/2}.
  \label{eqAppPNA}
\end{equation}
This result is used in \Sec{secEx} to calculate the mean area of a
freely jointed ring in the strong-swelling regime.

\bibliographystyle{unsrt}

\begin{thebibliography}{}
\bibitem{Butler1987} M.G. Brereton, C. Butler, J. Phys. A \textbf{20}, 3955 (1987).
\bibitem{Wiegel1988} 
D.C. Khandekar, F.W. Wiegel, J. Phys. A \textbf{21}, L563 (1988); J. Phys. France \textbf{50}, 263 (1989).
\bibitem{Duplantier1989} B. Duplantier, J. Phys. A \textbf{22}, 3033 (1989).
\bibitem{Duplantier1990} B. Duplantier, Phys. Rev. Lett. \textbf{64}, 493 (1990).
\bibitem{Cardy1994} J. Cardy, Phys. Rev. Lett. \textbf{72}, 1580 (1994).
\bibitem{Fisher1987} S. Leibler, R.R.P. Singh, M.E. Fisher, Phys. Rev. Lett. \textbf{59}, 1989 (
1987).
\bibitem{Fisher1990} A.C. Maggs, S. Leibler, M.E. Fisher, C.J. Camacho, Phys. Rev. A \textbf{42}
, 691 (1990).
\bibitem{Fisher1990_2} C.J. Camacho, M.E. Fisher, Phys. Rev. Lett. \textbf{65}, 9 (1990).
\bibitem{Rudnick1991} J. Rudnick, G. Gaspari, Science \textbf{252}, 422 (1991). 
\bibitem{Rudnick1993} G. Gaspari, J. Rudnick, A. Beldjenna, J. Phys. A \textbf{26}, 1 (1993).
\bibitem{Levinson1992} E. Levinson, Phys. Rev. A \textbf{45}, 3629 (1992).
\bibitem{Maritan1993} U.M.B. Marconi, A. Maritan, Phys. Rev. E \textbf{47}, 3795 (1993).
\bibitem{Gaspari1993} G. Gaspari, J. Rudnick, M. Fauver, J. Phys. A \textbf{26}, 15 (1993).
\bibitem{Haleva2006} E. Haleva, H. Diamant, Eur. Phys. J. E \textbf{19}, 461 (2006).
\bibitem{Vilanove1980} R. Vilanove, F. Rondelez, Phys. Rev. Lett. \textbf{45}, 1502 (1980).
\bibitem{Gavranovic2005} G.T. Gavranovic, J.M. Deutsch, G.G. Fuller, Macromolecules \textbf{38}, 6672 (2005).
\bibitem{Maier2000} B. Maier, J.O. Radler, Macromolecules \textbf{33}, 7185 (2000).
\bibitem{BenNaim2001} E. Ben-Naim, Z.A. Daya, P. Vorobieff, R.E. Ecke, Phys. Rev. Lett. \textbf{86}, 1414 (2001); R.E. Ecke, Z.A. Daya, M.K. Rivera, E. Ben-Naim, in \textit{Materials Research Society Symposium Proceedings \textbf{759}, 2003}, edited by S. Sen, M.L. Hunt, and A.J. Hurd (Warrendale, PA, 2003), p. 129.
\bibitem{BenNaim2006} M.B. Hastings, Z.A. Daya, E. Ben-Naim, R.E. Ecke, Phys. Rev. E \textbf{66}, 025102 (2002); Z.A. Daya. E. Ben-Naim, R.E. Ecke, e-print cond-mat/0603301.
\bibitem{PolymerPhysics2003} M. Rubinstein, R.H. Colby \textit{Polymer Physics} (Oxford University Press, Oxford, 2003).
\bibitem{Harris1966} R.A. Harris, J.E. Hearst, J. Chem. Phys. \textbf{44}, 2595 (1966); \textbf{46}, 398 (1967).
\bibitem{Thirumalai1995} B.Y. Ha, D. Thirumalai, J. Chem. Phys. \textbf{103}, 9408 (1995); e-print cond-mat/9709345.
\bibitem{Diamant2000} H. Diamant, D. Andelman, Phys. Rev. E \textbf{61}, 6740 (2000).
\bibitem{Gompper1997} G. Gompper, D.M. Kroll, J. Phys. Condens. Matter {\bf 9}, 8795 (1997).

% Format for Journal Reference \bibitem{RefJ} Author, Journal \textbf{Volume}, (year) page numbers.
% Format for books \bibitem{RefB} Author, \textit{Book title} (Publisher, place year) page numbers

\end{thebibliography}

\end{document}